%% file: main.tex
\documentclass[conference, 9pt]{IEEEtran}
\IEEEoverridecommandlockouts

\usepackage{verbatim}
\usepackage{amsmath}
\usepackage{amssymb}
\usepackage{booktabs} 
\usepackage{algorithm}
\usepackage{algorithmicx} 
\usepackage{algcompatible}
\usepackage{algpseudocode}
\usepackage{multirow}
\usepackage{color}
\usepackage{threeparttable}
\usepackage{pifont}
\usepackage{etaremune}
\usepackage{caption}
\usepackage{subcaption}
\usepackage[export]{adjustbox}
\usepackage{todonotes}
\usepackage{makecell}

\usepackage{enumitem}
\setlist{leftmargin=3.0mm}

\usepackage{soul}

\usepackage[english]{babel}
\usepackage[utf8]{inputenc}
\usepackage{url}
    
\usepackage{lipsum}

\newcommand\blfootnote[1]{%
  \begingroup
  \renewcommand\thefootnote{}\footnote{\scriptsize #1}%
  \addtocounter{footnote}{-1}%
  \endgroup
}
\makeatletter
\def\footnoterule{\relax%
  \kern-5pt
  \hbox to \columnwidth{\hfill\vrule width 0.9\columnwidth height 0.4pt\hfill}
  \kern 4.6pt}
\makeatother
    
\begin{document}

\title{
\huge
\vspace{-12pt}
\textsc{MeLoPPR:} Software/Hardware Co-design for \underline{Me}mory-efficient \underline{Lo}w-latency \underline{P}ersonalized \underline{P}age\underline{R}ank
}

\author{Lixiang Li$^{1}$, Yao Chen$^2$, Zacharie Zirnheld$^3$, Pan Li$^3$, and Cong Hao$^4$ \\
\small
$^1$Dalhousie University, Canada, $^2$Advanced Digital Sciences Centre (ADSC), Singapore, \\
$^3$Purdue University, USA, $^4$Georgia Institute of Technology, USA
}

\IEEEoverridecommandlockouts
\IEEEpubid{\makebox[\columnwidth]{978-1-6654-3274-0/21/\$31.00 \textcopyright 2021 IEEE\hfill} \hspace{\columnsep}\makebox[\columnwidth]{ }}

\maketitle

\begin{abstract}

Personalized PageRank (PPR) is a graph algorithm that evaluates the importance of the surrounding nodes from a source node. Widely used in social network related applications such as recommender systems, PPR requires real-time responses (latency) for a better user experience.
Existing works either focus on algorithmic optimization for improving precision while neglecting hardware implementations or focus on distributed \textit{global graph processing} on large-scale systems for improving throughput rather than response time. 
Optimizing low-latency local PPR algorithm with a tight memory budget on edge devices remains unexplored. In this work, we propose a \underline{me}mory-efficient, \underline{lo}w-latency PPR solution, namely MeLoPPR, with largely reduced memory requirement and a flexible trade-off between latency and precision. MeLoPPR is composed of stage decomposition and linear decomposition and exploits the node score sparsity:
Through stage and linear decomposition, MeLoPPR breaks the computation on a large graph into a set of smaller sub-graphs, that significantly saves the computation memory; Through sparsity exploitation, MeLoPPR selectively chooses the sub-graphs that contribute the most to the precision to reduce the required computation.
In addition, through software/hardware co-design, we propose a hardware implementation on a hybrid CPU and FPGA accelerating platform, that further speeds up the sub-graph computation. We evaluate the proposed MeLoPPR on memory-constrained devices including a personal laptop and Xilinx Kintex-7 KC705 FPGA using six real-world graphs. First, MeLoPPR demonstrates significant memory saving by $1.5\times \sim 13.4\times$ on CPU and $73\times \sim 8699\times$ on FPGA.
Second, MeLoPPR allows flexible trade-offs between precision and execution time: when the precision is $80\%$, the speedup on CPU is up to $15\times$ and up to $707\times$ on FPGA; when the precision is around $90\%$, the speedup is up to $70\times$ on FPGA.
\end{abstract}

\input{sec1.tex}
\input{sec2.tex}

\input{sec3.tex}

\input{sec4.tex}

\input{sec5.tex}

\input{sec6.tex}

\input{sec7.tex}

\section*{Acknowledgement}
This project is partially supported by the National Research Foundation, Prime Minister's Office, Singapore under its Campus for Research Excellence and Technological Enterprise (CREATE) programme.

\bibliographystyle{unsrt}
\footnotesize
\bibliography{ref}

\end{document}

%% file: sec1.tex
\section{Introduction}
\vspace{-6pt}

\blfootnote{Code available at: \url{https://github.com/sharc-lab/MeLoPPR}}

Personalized PageRank (PPR) is a basic algorithm widely used by many web applications, such as recommender system, social network community detection, etc~\cite{jeh2003scaling}. 
On a graph, given a source node, PPR evaluates the importance and relevance of the surrounding nodes with respect to the source node, usually by identifying the top-$k$ nodes with the highest PPR scores.
PPR is especially useful in commercial applications such as who-to-follow recommendations of Twitter and Facebook, and related product recommendations of Amazon.
Therefore, efficient PPR computation can greatly improve user experience (e.g., faster response time).
However, their computations are very challenging in real-world applications because of: 1) Extremely large graph size. Many real-world graphs have trillions of edges and nodes with sizes of up to tens of Giga bytes. It's impossible to load the whole graph into memory during computation, thus introducing considerable data movement overhead; 2) Irregular memory access and poor data locality. 
Most graph algorithms, especially PPR, introduce significant glitches in memory access.

These challenges usually result in large memory requirement and/or long computational latency introduced by data swapping in and out.
In most cases, the computational latency, i.e., response time, is an important objective of a PPR server, that whenever a seed node is queried, a response with top-$k$ most related nodes is expected as quickly as possible.
Although the response time can be largely reduced by assuming sufficient memory to store the graph on a large-scale distributed server system, it is not always realistic not only because of the huge graph size, but also because of the randomness of the seed node to be queried.
On the other hand, when the PPR computation must be conducted on memory constrained devices such as personal laptops or edge devices (e.g., for privacy protecting), the query latency will be largely and further deteriorated.

Addressing these challenges and improving either computation efficiency or precision attracts researchers from both software and hardware communities. At the software side, some existing works focus on algorithm optimization for improving PPR computation \textit{precision} \cite{fujiwara2012efficient, wang2019efficient}, but neglect hardware implementations. Some works propose efficient PPR algorithms on large scale or distributed server systems, such as PowerWalk \cite{liu2016powerwalk}, Fast-PPR \cite{lofgren2014fast}, TopPPR \cite{wei2018topppr} and Fora \cite{wang2017fora}, but still requires a significant amount of memory (up to Giga bytes) and/or heavy pre-processing (up to hours). At the hardware side, most power graph processing systems, such as GraphH \cite{dai2018graphh}, Blogel \cite{yan2014blogel} and Giraph++ \cite{tian2013think}, are targeting general-purpose \textit{global} graph algorithms such as global pagerank, connected components, etc., for improving overall \textit{throughput} rather than latency.

So far, flexible and well-balanced solutions between the \textit{memory requirement}, \textit{latency}, and \textit{precision} are still missing.
Especially, given the recent paradigm shifting from server computation to edge computation, it is necessary that the PPR being executed on memory-constrained devices but still with satisfying latency and precision.
In addition, the gap between algorithm optimization and hardware implementation has to be closed.
Driven by these emerging needs, in this work, we propose \textbf{MeLoPPR}, a \underline{\textbf{me}}mory-efficient, \underline{\textbf{lo}}w-latency \underline{\textbf{PPR}} software/hardware co-design solution, that can flexibly balance between different objectives and constraints, especially with tight memory budget.
We summarize our contributions as follows:
\begin{itemize}
    \item {
    We are the first to propose a hardware-friendly multi-stage PPR algorithm, namely MeLoPPR, with significantly reduced memory requirements through \textit{stage decomposition and linear decomposition} of large graphs. The decomposed sub-graphs also open up opportunities for parallel computing.
    }
    \item {
    MeLoPPR exploits the \textit{sparsity of the PPR vectors} to largely reduce the required computation, providing flexible trade-offs between the computation latency and PPR precision.
    }
    \item {
    Through software/hardware co-design, we propose a dedicated FPGA accelerator of MeLoPPR, greatly shortening the latency by cooperating with a host CPU. We also propose hardware-aware optimizations, such as localized score aggregation, to further reduce the overall latency and memory requirement.
    }
    \item {
    We evaluate the proposed MeLoPPR algorithm and its accelerator on a personal laptop and a Xilinx Kintex-7 FPGA, and demonstrate that MeLoPPR achieves remarkable memory saving by $1.5\times \sim 13.4\times$ on CPU and $73\times \sim 8699\times$ on FPGA, respectively.
    Equipped with the flexibility for computation and latency trade-off, MeLoPPR achieves up to $15\times$ speedup on CPU under a $80\%$ precision and up to $70\times$ speedup on FPGA under a $90\%$ precision.
    }
\end{itemize}


%% file: sec2.tex
\section{Preliminaries \label{sec:prelim}}


Assume $G = (V, E)$ is a simple, undirected graph. We define the size of $G$ as $|V|+|E|$ and later use $O(G)$ to denote $O(|V|+|E|)$.

 \textbf{$\alpha$-decay Random Walk ($\alpha$-RW)}. 
 Given a source (seed) node $s \in V$, a random walk (RW) starts from $s$ and proceeds to a random neighbor of the current node. 
$\alpha$-RW is RW paired with a decay factor $\alpha$.  
At each step, 
$\alpha$-RW either terminates at the current node with $1-\alpha$ probability, or proceeds to a randomly selected out-neighbor of the current node. $\alpha$-RW can be used to compute \emph{PageRank} \cite{page1999pagerank} where the source node $s$ is chosen randomly.




\textbf{Personalized PageRank (PPR)}~\cite{jeh2003scaling}. 
PPR is a variant of PageRank with only one fixed source node $s$. Given another node $v$, we use $\pi(v)$ to denote its PPR score, which is the probability that $\alpha$-RW stars from $s$ and terminates at $v$. Typically, in practice, the exact value $\pi(v)$ is less important. Instead, the top-$k$ nodes with the highest $\pi(v)$ values are needed. Hence, in this work, we focus on detecting these top-$k$ nodes. We denote the accurate (ground-truth) set of top-$k$ nodes from $s$ as $T(s, k)$,
and the set given by our method as $\hat{T}(s, k)$.

\textbf{Graph Diffusion}.
PPR can be formulated and solved by graph diffusion \cite{park2019survey}.
Denote the degree of a node $v\in V$ by $d_v$, the associated adjacency matrix of $G$ by $A$, and the diagonal matrix of degrees $(D_{ii} = d_i)$ by $D$. Let $W = AD^{-1}$ be the \textit{random walk transition} matrix. 
In this work, we focus on $\alpha$-RW of maximum $L$ steps. Then, given an initial vector $S_0\in\mathbb{R}^{|V|}$, graph diffusion computes $S_l$ ($1\leq l \leq L$) recursively as:
\begin{equation}
\begin{split}
 {S_{l+1}} & = (1-\alpha) \cdot {S_{0}} + \alpha \cdot W \cdot {S_l} \\
 {S_L} 
  & = (1-\alpha) \sum_{k=0}^{L-1} \alpha^k W^k S_0 + \alpha^L W^L S_0.
 \end{split}
 \label{eq:diffusion-computation}
\end{equation}
 We denote this graph diffusion as $S_l = \mathcal{GD}^{(l)}(S_0)$ for $0\leq l \leq L$. In PPR, the initial vector $S_0$ are all zeros except for the source node $s$ with a value one. The final output $S_L$ gives the PPR scores with such an $S_0$ initialization.  
 
\begin{figure}
    \centering
    \includegraphics[width=0.5\textwidth]{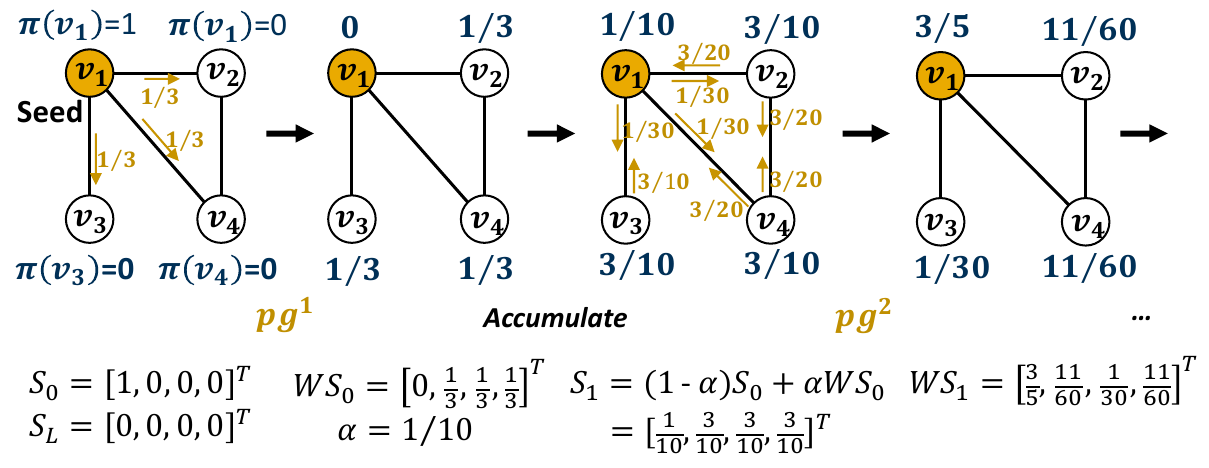}
    \caption{The propagations ($pg^1$, $pg^2$, etc.) and accumulations within graph diffusion $\mathcal{GD}^{(L)}(S_0)$ where $v_1$ is the seed node.}
    \label{fig:gd-propagation}
    \vspace{-8pt}
\end{figure}

Each iteration in Eq.~\eqref{eq:diffusion-computation} involves two steps, \textit{propagation} (denoted as $pg^1$, $pg^2$, etc.) and \textit{accumulation}. Fig.~\ref{fig:gd-propagation} shows an example of the propagation and accumulation.
One may use the graph diffusion operation on $G$ to obtain $T(s, k)$ as follows:
\begin{equation}
     T(s, k) \leftarrow \mathcal{R} (S_L, k), \quad \text{where} \;S_L = \mathcal{GD}^{(L)}(S_0).
\label{eq:diffusion-notation}
\end{equation}
Here $\mathcal{R}$ selects $k$ nodes in the descending order based on their PPR scores in $S_L$. Our goal is to find $\hat{T}(s, k)$ that approximates $T(s, k)$.

\textbf{Measurement}.
To measure how many nodes in the accurate top-$k$ node set $T$ are correctly found, we use \textit{precision}, denoted as $Prec(s, k)$, to quantify the accuracy of an approximated node set $\hat{T}$, computed as
$Prec(s, k) = |\{v| v\in \hat{T}(s, k) \land v\in T(s, k) \}| / k$.


%% file: sec3.tex
\section{motivation and related works}
\label{sec:motivation}

\begin{figure*}
\vspace{-16pt}
    \centering
    \includegraphics[width=0.92\textwidth]{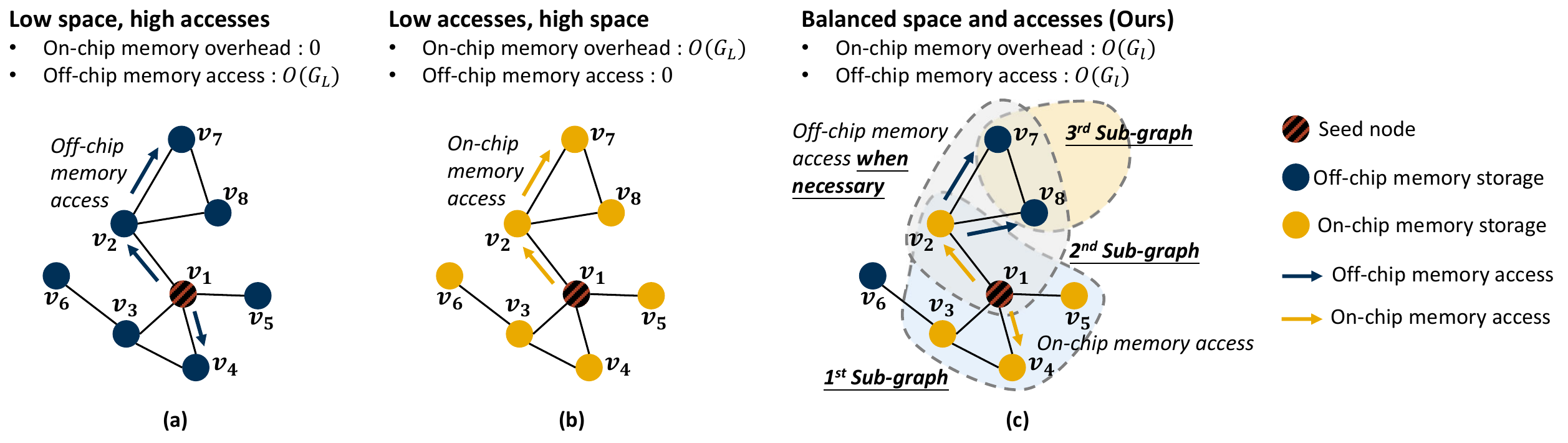}
    \vspace{-6pt}
    \caption{(a) Methods with low on-chip memory requirement and high off-chip access, such as MC random walk. (b) Methods with low off-chip memory access and large on-chip requirement (the whole related sub-graph is loaded into memory). (c) Our proposed multi-stage MeLoPPR with balanced on-chip memory and off-chip access, by adaptively loading only the necessary sub-graphs.}
    \label{fig:motivation}
    \vspace{-8pt}
\end{figure*}

    

\textbf{Software-emphasis works}. 
Given the importance of efficient PPR computing, a great amount of effort has been done to improve ranking accuracy or to theoretically reduce computation complexity.
Fujiwara et. al~\cite{fujiwara2012efficient} propose to compute node relevance from sparse matrices with theoretical exactness guarantee.
Fast-PPR \cite{lofgren2014fast} proposes a method with a theoretical running-time guarantee of $O(\sqrt{d / \delta})$ where $d$ is the average in-degree of the nodes, and $\delta$ is the error tolerance.
Fora~\cite{wang2017fora} proposes index-free and index-based algorithms for approximate PPR computation, with rigorous guarantees on result quality. TopPPR~\cite{wei2018topppr} focuses on exact top-$k$ PPR queries and ensures a precision of $\rho$ with $1-1/n$ probability.
Despite these great achievements, they still have considerable memory requirements. For instance, Fora~\cite{wang2017fora} introduces up to 81.5 GB memory overhead, while TopPPR~\cite{wei2018topppr} assumes that the entire graph fits into 96 GB main memory.
Another observation goes to the gap between algorithmic analysis and actual hardware implementation: a theoretical reduction in memory or latency does not always lead to better hardware performance.
For example, it is stated that the space overhead of classic Monte Carlo (MC) random walk is zero \cite{wei2018topppr}; However, the reduced on-chip space will result in a significant off-chip data access overhead, as shown in Fig.~\ref{fig:motivation} (a). 

Therefore, we intend to close the gap between the software algorithm and hardware implementation, which \textit{significantly reduces the on-chip memory requirement, with an explicit awareness of the on-chip memory space and the off-chip memory access}, as illustrated in Fig.~\ref{fig:motivation} (c).
More details are in Section~\ref{sec:proposed-algorithm}.

\textbf{Hardware-emphasis works}.
Another category of existing works is large-scale graph processing system, such as GraphH \cite{dai2018graphh}, Blogel \cite{yan2014blogel}, Giraph++~\cite{tian2013think}, etc. They aim at improving graph process \textit{throughput} when running \textit{global} algorithms such as PageRank~\cite{page1999pagerank} and connected components; most of them require graph pre-processing, which is not suitable for \textit{local} algorithms such as PPR that requires short \textit{latency}.
Therefore, we propose a dedicated hardware accelerator for PPR. More details are in Section~\ref{sec:co-design}.


%% file: sec4.tex
\section{proposed multi-stage MeLoPPR} \label{sec:proposed-algorithm}

In this section we introduce the proposed MeLoPPR from algorithm aspect.
We first intuitively explain the overall idea in Sec.~\ref{sec:overall-idea}, and then provide mathematical formulations in the following sections.

\subsection{Overall Idea}
\label{sec:overall-idea}
To efficiently compute the local PPR from a seed node $s$ within maximum length of $L$, the ideal method is to extract a sub-graph from $s$ with a BFS of depth $L$ and load the sub-graph into on-chip memory before computation, as illustrated in Fig.~\ref{fig:motivation} (b).
However, the sub-graph size and BFS time grow exponentially with $L$, making it unrealistic with a tight memory budget and limited latency requirement.
Therefore, as illustrated in Fig.~\ref{fig:motivation} (c), MeLoPPR adaptively breaks the large graph into a set of smaller sub-graphs that can entirely fit into the on-chip memory.
MeLoPPR has two primary features:
\begin{enumerate}
    \item {
    \textit{\textbf{It achieves memory-efficiency through stage decomposition and linear decomposition}}. When a seed node $s$ is queried ($v_1$ in this example), we apply \textit{stage decomposition} (formulations in Section~\ref{sec:stage-decomp}). In stage-one, we first run a BFS of depth $l$ to extract a smaller sub-graph denoted as $G_{l}(s)$, and then execute a graph diffusion with $l$-iterations on $G_{l}(s)$. $l$ is smaller than $L$ so that $G_l(s)$ can be loaded into the on-chip memory.
    In stage-two and afterwards, we continue to extract sub-graphs from the nodes in the first sub-graph, called \textit{next-stage nodes}, and apply graph diffusion to compute and calibrate the PPR scores. 
    This is called \textit{linear decomposition} (formulations in Section~\ref{sec:linear-decomp}).
    In the case of Fig.~\ref{fig:motivation} (c), the sub-graph from $v_2$ and the sub-graph from $v_7$ are extracted in the linear decomposition.
    In this way, by decomposing one large graph diffusion into consecutive multiple-step diffusion of smaller graphs, the required on-chip memory can be greatly reduced.
    
    }
    \item {
    \textit{\textbf{It achieves low-latency, as well as latency-precision trade-off, due to the sparsity of the PPR vector}}.
    Generally, the more next-stage nodes are selected, the longer computation latency and the higher precision are expected.
    Fortunately, the PPR vector after the graph diffusion is very sparse, i.e., a majority of the nodes have PPR scores close to zero.
    Therefore, MeLoPPR needs to choose only a small amount of next-stage nodes, which significantly reduces the overall computation latency.
    Formal formulation and more details are provided in Section~\ref{sec:linear-decomp} and~\ref{sec:sparsity}.
    }
\end{enumerate}

\subsection{Stage Decomposition}
\label{sec:stage-decomp}

\begin{figure*}
\vspace{-16pt}
    \centering
    \includegraphics[width=0.95\textwidth]{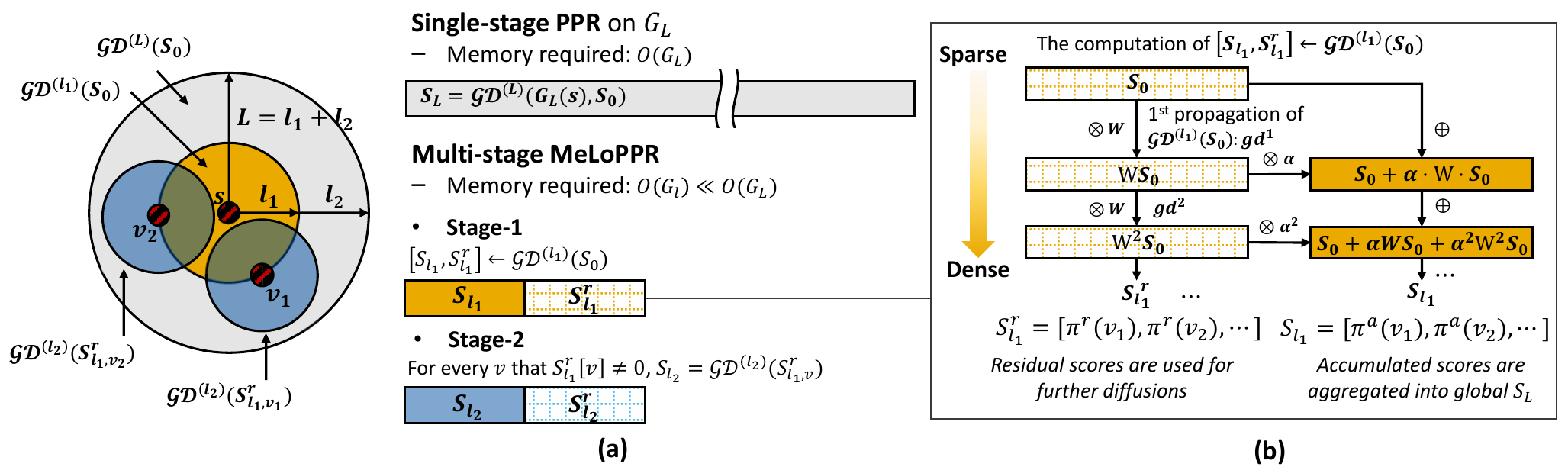}
    \caption{(a) An illustration of memory requirements of the single-stage PPR (gray circle) and the multi-stage MeLoPPR (yellow and blue circles). (b) The computation flow within one graph diffusion to obtain the accumulated scores ($\pi^a$) and the residual scores ($\pi^r$).}
    \label{fig:GD_hierarchy}
\vspace{-8pt}
\end{figure*}

Without loss of generality, we divide the graph diffusion of length $L$ into $L = l_1 + l_2$, which can be easily extended to more terms. Eq.~\ref{eq:diffusion-computation} can be expressed as:
\begin{equation}
\footnotesize
\mathcal{GD}^{(L)}(S_0) = (1-\alpha) \sum_{k = 0}^{l_1 + l_2 - 1} \alpha^k  W^k  S_0 + \alpha^L W^L S_0
\label{eq:gd_split-1}
\end{equation}
The first term in Eq.~\ref{eq:gd_split-1} can be expressed as:
\begin{equation}
\footnotesize
\begin{split}
& (1-\alpha) \sum_{k = 0}^{l_1 + l_2 - 1} \alpha^k  W^k  S_0 \\
& =  \mathcal{GD}^{(l_1)} (S_0) - \alpha^{l_1} W^{l_1} S_0 + (1-\alpha) \sum_{k = l_1}^{l_1 + l_2-1} \alpha^k W^k  S_0  \\
\end{split}
\label{eq:gd_split-2}
\end{equation}
Similarly, the last term in Eq.~\ref{eq:gd_split-2} is expressed as:
\begin{equation}
\footnotesize
\begin{split}
& (1-\alpha) \sum_{k = l_1}^{l_1 + l_2 - 1} \alpha^k  W^k  S_0 = (1-\alpha) \cdot \alpha^{l_1}  \sum_{k = 0}^{l_2-1}  \alpha^{k}  W^{k}  ( W^{l_1}  S_0) \\
& = \alpha^{l_1}  \mathcal{GD}^{(l_2)} ( W^{l_1}  S_0 ) - \alpha^{l_1} \alpha^{l_2} W^{l_2} (W^{l_1}  S_0) \\
& = \alpha^{l_1}  \mathcal{GD}^{(l_2)} ( W^{l_1}  S_0 ) - \alpha^{L} W^{L} S_0 \\
\end{split}
\label{eq:gd_split-3}
\end{equation}
Summing up Eq.~\ref{eq:gd_split-3}, Eq.~\ref{eq:gd_split-2}, and Eq.~\ref{eq:gd_split-1}:
\begin{equation}
\footnotesize
\begin{split}
& \mathcal{GD}^{(L)}(S_0) = \mathcal{GD}^{(l_1)} (S_0) + \alpha^{l_1}  \mathcal{GD}^{(l_2)} ( W^{l_1}  S_0 ) - \alpha^{l_1} W^{l_1}  S_0\\
\end{split}
\label{eq:diffusion-split}
\end{equation}

Easy to know,
$\mathcal{GD}^{(l_1)} (S_0)$ is the graph diffusion with $l_1$ iterations, starting with the initial vector $S_0$ and outputting the vector $S_{l_1}$;
$\mathcal{GD}^{(l_2)} ( W^{l_1} S_0)$ is the graph diffusion with $l_2$ iterations starting with the initial vector $ W^{l_1} S_0 $.
Since $W^{l_1} S_0 \neq S_{l_1}$, 
we denote $W^{l_1} S_0$ as $S_{l_1}^r$ to distinguish from $S_{l_1}$ and maintain $S_{l_1}$ and $S_{l_1}^r$ separately during graph diffusion.
Fig.~\ref{fig:GD_hierarchy} (b) illustrates the computation flow for obtaining $S_{l_1}$ and $S_{l_1}^r$.

In this way, a graph diffusion with total $l_1+l_2$ iterations can be decomposed as two stages of consecutive graph diffusion operations.

\subsection{Linear Decomposition}
\label{sec:linear-decomp}

Applying stage decomposition alone, however, does not save the required memory, as illustrated in Fig.~\ref{fig:GD_hierarchy} (a).
The largest gray sub-graph represents the original graph diffusion with depth $L$; the required memory is $|S_L|=O(G_{L}(s))$, proportional to the size of the gray sub-graph, $G_{L}(s)$.
The yellow sub-graph represents the stage-one graph diffusion with depth $l_1$.
The stage-two graph diffusion with depth $l_2$, as defined in Eq.~\eqref{eq:diffusion-split}, can be seen as propagating from the yellow sub-graph to the gray sub-graph. Doing so, however, the required memory is still proportional to the size of $G_{L}(s)$.


Fortunately, graph diffusion has a \textit{linearity property} that can split one diffusion into multiple ones.
We write $S_{l_1}^r = \sum_{v\in V}S_{l_1,v}^r$ where $S_{l_1,v}^r$ is a vector that zeros all the components in $S_{l_1}^r$ except the one corresponding to $v$. Specifically, if $S_{l_1}^r = 
(S_{l_1}^r[v_1], S_{l_1}^r[v_2], S_{l_1}^r[v_3], \cdots)^T$,
then $S_{l_1,v_1}^r = (S_{l_1}^r[v_1], 0, 0, \cdots)^T$,
$S_{l_1,v_2}^r = (0, S_{l_1}^r[v_2], 0, \cdots)^T$, etc.
Therefore, $\mathcal{GD}^{(l_2)} (S_{l_1}^r)$ can be written into a combination of $\mathcal{GD}^{(l_2)} (S_{l_1,v}^r)$ over non-zero $S_{l_1,v}^r$'s. Note that the nodes $v$ with non-zero $S_{l_1,v}^r$'s are in  $G_{l_1}(s)$. Hence,  
\begin{equation}
    \mathcal{GD}^{(l_2)}(S_{l_1}^r) = \sum_{v\in G_{l_1}(s)} \mathcal{GD}^{(l_2)} (S_{l_1,v}^r)
    \label{eq:linear-decompose}
\end{equation}
As shown in Fig.~\ref{fig:GD_hierarchy} (a), the second-stage linear decomposition is represented by the blue graphs; the required memory is proportional to the sub-graph size of $G_{l_2}(v)$, which is much smaller than that of the original sub-graph $G_{L}(s)$.

Substituting Eq.~\eqref{eq:linear-decompose} into Eq.~\eqref{eq:diffusion-split}, the single-stage PPR is decomposed into multi-stage MeLoPPR as below:
\begin{equation}
\begin{split}
\footnotesize
\mathcal{GD}^{(l_1 + l_2)}(S_0) & = \mathcal{GD}^{(l_1)} (S_0) - \alpha^{l_1}   S_{l_1}^r  \\
& + \alpha^{l_1} \sum_{v\in G_{l_1}(s)} \mathcal{GD}^{(l_2)} (S_{l_1,v}^r)
\end{split}
\label{eq:multi-stage-PPR}
\end{equation}

\subsection{Sparsity of the PPR Vector}
\label{sec:sparsity}


Although stage and linear decomposition saves the memory requirement and the BFS time,
according to Eq.~\ref{eq:multi-stage-PPR}, an accurate PPR computation requires that graph diffusion to be executed on each node $v\in G_{l_1}(s)$, which may increase the computation latency.
Luckily, we observe that the PPR vector after the graph diffusion is highly sparse. In the bottom figure in Fig.~\ref{fig:sparsity-and-accuracy},
we show the normalized PPR score distribution in log scale, obtained after the stage-one PPR on a real-world graph~\cite{snapnets}:
only less than $1\%$ of the total nodes inside $G_{l_1}(s)$ have relatively large PPR scores, while more than $90\%$ of the nodes have close-to-zero scores.
Such a great sparsity of the PPR vector allows our proposed MeLoPPR to focus only on a small subset of the nodes, called \textit{next-stage nodes}, while still achieves a descent precision.
Intuitively, for a node $v$, the larger the residual score $S_{l_1}^r[v]$ is, the more desirable $v$ shall be selected for the next-stage computation;
therefore, the next-stage nodes are selected in the descending order based on their residual scores $S_{l_1}^r[v]$. 



%% file: sec5.tex
\section{hardware implementation}
\label{sec:co-design}

On top of our proposed MeLoPPR algorithm, we propose a dedicated hardware accelerator for MeLoPPR through multiple software/hardware co-optimization techniques on a hybrid CPU+FPGA System-on-Chip platform.
The overall implementation is shown in Fig.~\ref{fig:overall_FPGA_CPU}.
The processing system (PS) refers to the CPU execution, and the programming logic (PL) refers to the FPGA execution. 
First, the CPU works as the overall controller: it prepares the sub-graph through BFS, reorganizes the sub-graph into a list of nodes with neighbors, communicates with FPGA, and collects the intermediate and final results.
Second, the FPGA works as an off-loading device for graph diffusion computation, given its massive parallelism capability.
The design challenges include:
1) To effectively exploit parallelism on FPGA to accelerate one graph diffusion operation;
2) To effectively reduce the data transfer overhead between CPU and FPGA.
In the following sections, we discuss the proposed solutions to these challenges.

\subsection{FPGA Implementation for Diffusion Acceleration}

\begin{figure}
    \centering
    \includegraphics[width=0.49\textwidth]{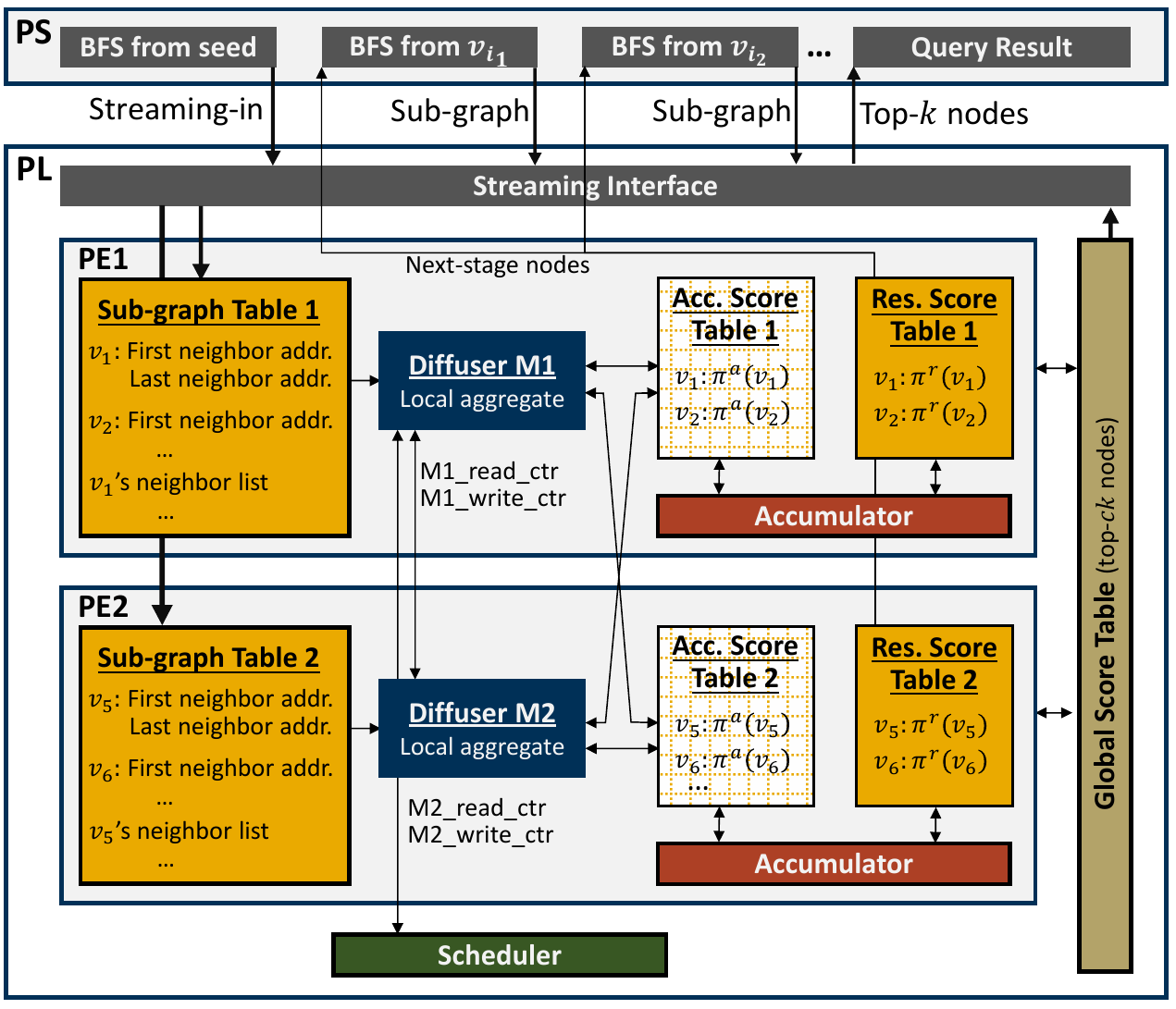}
    \caption{The overall block diagram of the proposed CPU+FPGA implementation of MeLoPPR.}
    \label{fig:overall_FPGA_CPU}
\end{figure}

To compute the diffusion on one sub-graph, we design a processing element (PE) that composes of five components, as shown in Fig.~\ref{fig:overall_FPGA_CPU}:
1) a sub-graph table, that store the node's neighbors and their address information;
2) a local accumulated score table (Acc. Score), that stores the current accumulated PPR scores $\pi^a(v)$ in each iteration of the nodes;
3) a local residual score table (Res. Score), that stores the current residual scores $\pi^r(v)$ in each iteration (Sec.~\ref{sec:linear-decomp});
4) a diffuser module, that reads the nodes from the sub-graph tables, fetches the scores from the score table, computes the current propagation, and writes back the updated scores to the score tables;
5) an accumulator, that computes the node scores $\pi^a(v)$ and $\pi^{r}(v)$ following Fig.~\ref{fig:GD_hierarchy} (b).
There is also a global score table, that stores the PPR scores of the top $c\cdot k$ nodes ($c$ specified in Sec.~\ref{sec:global-graph-aggregation}).

To improve graph diffusion efficiency, parallel executions must be enabled. 
For parallelism $P$, $P$ PEs are instantiated.
In Fig.~\ref{fig:overall_FPGA_CPU} we demonstrate the parallelism of 2 with two PEs for simplicity, and more scalability studies are provided in Sec.~\ref{sec:fpga-scalability}.
Each diffuser module reads from the sub-graph table within its own PE and writes to all the local score tables.
Thus, a scheduler must be created to resolve the read and write conflicts when different diffusers are to access the same local score table. 

Originally, the initial values in the PPR vector are one and zeros, and then become fractions represented by high-precision floating point, which is highly inefficient on FPGA.
Thus, we use 32-bit integers to represent the values in PPR vectors by assigning a large enough integer $Max$ to the seed node, where $Max= d \times |G_{L}(s)|$. During diffusion, all the computations are done in integer format. The multiplications with fractional coefficient $\alpha$ is represented as $\alpha \approx \alpha_p / \alpha_q$, where $\alpha_p$ is a 16-bit integer and $\alpha_q=2^q$, so the division by $\alpha_q$ is implemented as a $q$-bit shifting. 
Comparing the top-$k$ PPR precision between floating and integer computation, it shows that when $d$ equals the average degree of $G_{L}(s)$, the precision loss is less than $4\%$; when $d$ equals the maximum degree of $G_{L}(s)$, the precision loss is less than $0.001\%$.
In the final experiments we let $d$ to be half of the maximum degree and $q=10$.

\vspace{-4pt}
\subsection{Data Transfer Reduction}
\label{sec:global-graph-aggregation}

After each graph diffusion on a sub-graph, the output PPR vector must be aggregated into the global vector $S_L$ by summation, based on Eq.~\eqref{eq:multi-stage-PPR}.
Such score aggregation introduces large overhead because: 1) transferring the obtained vector after each graph diffusion from FPGA to CPU increases the overall latency;
2) maintaining the vector of $S_L$ also requires a large memory size of $O(G_{L}(s))$.
Since PPR only requires top-$k$ ranking nodes, it is unnecessary to keep the entire $S_L$.
Therefore, we maintain only a fixed-length global score table with size $c \cdot k$.
The global score table is maintained by FPGA in BRAM to avoid sending back the node scores to CPU after each sub-graph diffusion.
After all the diffusions, only the top-$k$ ranking nodes are sent back to CPU as the final query result.
The experiments show that when $c > 8$, the precision loss is less than $0.2\%$; and when $c < 4$, the precision loss is larger than $3\%$.
We let $c=10$ for final FPGA implementation, which significantly saves FPGA on-chip memory and reduces FPGA-CPU data transfer.


%% file: sec6.tex
\vspace{-4pt}
\section{experimental results}
\label{sec:experiments}

\begin{figure}
\vspace{-4pt}
    \centering
    \includegraphics[width=0.45\textwidth]{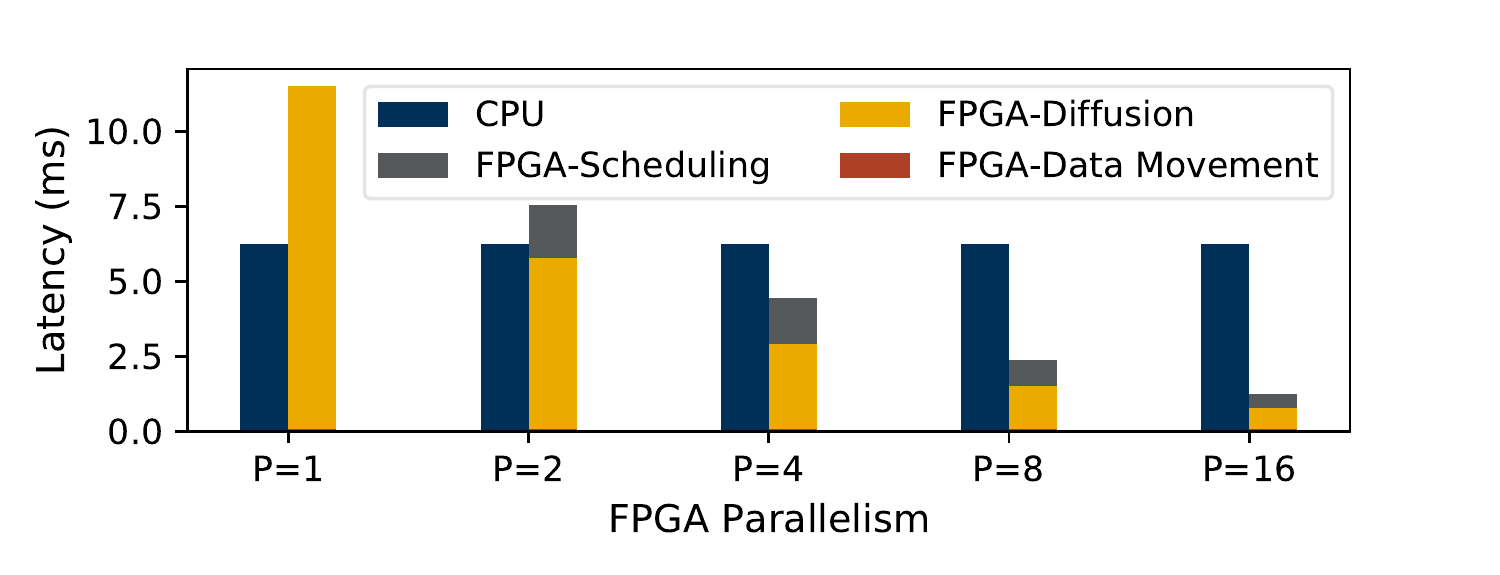}
    \vspace{-6pt}
    \caption{FPGA scalability with increased parallelism $P$.}
    \label{fig:fpga-scalability}
\end{figure}

\begin{table}[]
    \centering
    \begin{tabular}{c|c |c|c|c|c}
    \Xhline{2\arrayrulewidth}
    Resource & $P$=1 & $P$=2 & $P$=4 & $P$=8 & $P$=16 \\\hline
    LUTs    & 0.9\% & 3.1\% & 8.9\% & 21.8\% & 70.6\%  \\
    BRAM    & 4.8\% & 9.9\% & 19.2\% & 36.1\% & 72.8\% \\
    \Xhline{2\arrayrulewidth}
    \end{tabular}
    \caption{FPGA resource utilization under different parallelism $P$. The DSP usage is under $0.1\%$ since the division operations are implemented using logic.}
    \label{tab:FPGA-resource}
    \vspace{-8pt}
\end{table}

\begin{figure}[t]
    \centering
    \includegraphics[width=0.49\textwidth]{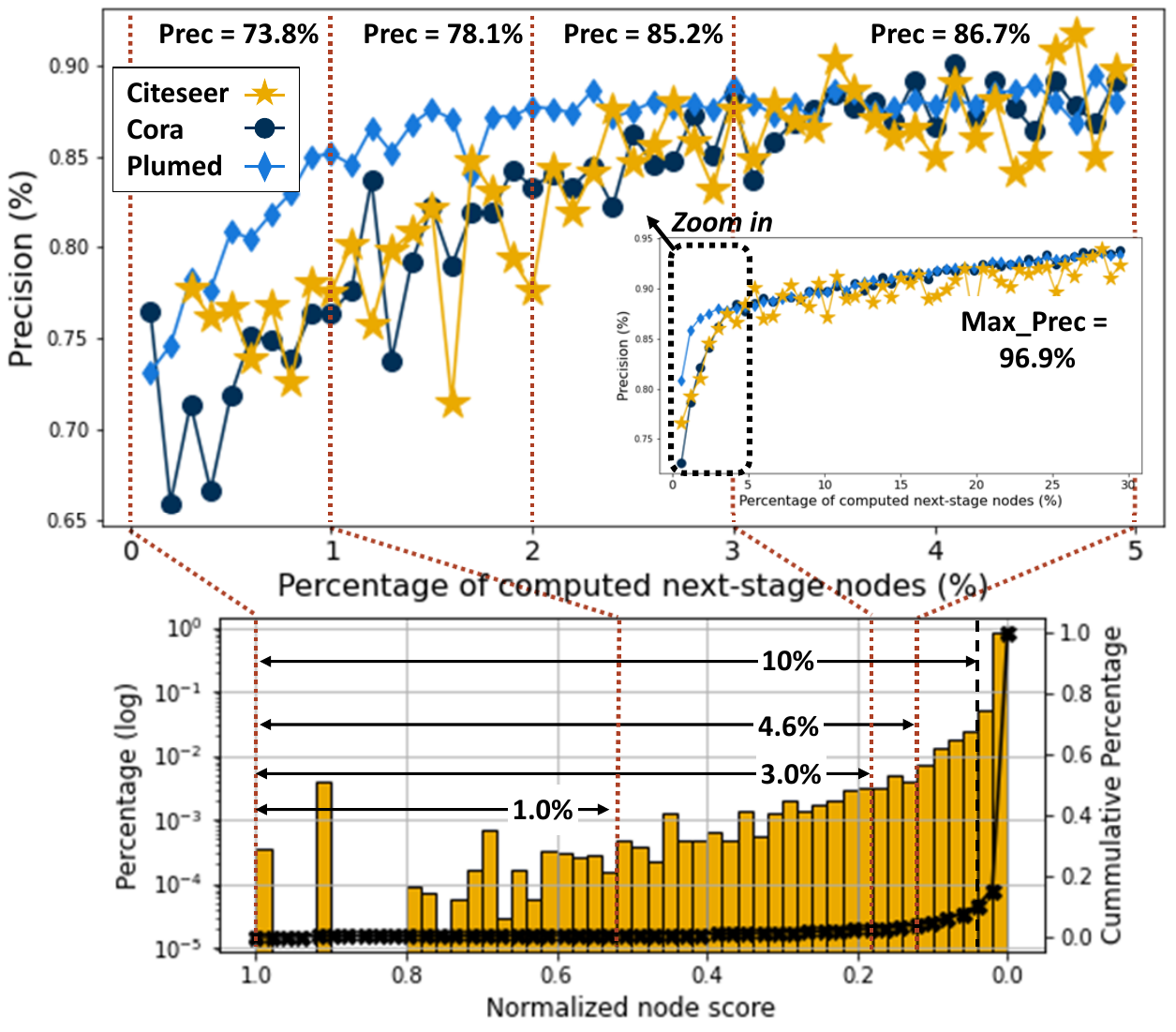}
    \caption{Exploiting the sparsity in PPR vector, the precision can reach more than $80\%$ using only $2\%$ of nodes. Using $5\%$ nodes, the precision can reach $96\%$.}
    \label{fig:sparsity-and-accuracy}
    \vspace{-8pt}
\end{figure}

\begin{table*}[]
\vspace{-8pt}
    \centering
    \footnotesize
    \setlength{\tabcolsep}{7pt}
    \begin{tabular}{c | c | c c c | c c c}
    \Xhline{2\arrayrulewidth}
    \textbf{Graph} & \textbf{LocalPPR-CPU} & \multicolumn{3}{c|}{\textbf{MeLoPPR-CPU (proposed)}} & \multicolumn{3}{c}{\textbf{MeLoPPR-FPGA (proposed)}} \\
    \cline{2-8}
    & Memory (MB) & Memory (MB) & Reduction & Avg. Red. & Memory (MB) & Reduction & Avg. Red. \\

\hline
\textbf{G1} &
0.005 $\sim$ 1.262 & 0.008 $\sim$ 0.723 & 0.55$\times$ $\sim$ 13.06$\times$ & 1.51$\times$ & 0.000 $\sim$ 0.021 & 23.97$\times$ $\sim$ 523.60$\times$ & 73.64$\times$ \\ 
\hline
\textbf{G2} &
0.005 $\sim$ 2.520 & 0.009 $\sim$ 1.427 & 0.83$\times$ $\sim$ 22.84$\times$ & 4.18$\times$ & 0.000 $\sim$ 0.037 & 45.20$\times$ $\sim$ 1173.96$\times$ & 214.58$\times$ \\ 
\hline
\textbf{G3} &
0.020 $\sim$ 20.727 & 0.028 $\sim$ 8.236 & 0.96$\times$ $\sim$ 17.21$\times$ & 6.43$\times$ & 0.000 $\sim$ 0.197 & 50.84$\times$ $\sim$ 3121.19$\times$ & 389.83$\times$ \\ 
\hline
\textbf{G4} &
0.012 $\sim$ 65.604 & 0.019 $\sim$ 5.264 & 0.84$\times$ $\sim$ 28.35$\times$ & 9.46$\times$ & 0.000 $\sim$ 0.134 & 28.40$\times$ $\sim$ 4785.45$\times$ & 595.55$\times$ \\ 
\hline
\textbf{G5} &
0.101 $\sim$ 320.490 & 0.032 $\sim$ 60.998 & 0.99$\times$ $\sim$ 21.57$\times$ & 13.43$\times$ & 0.000 $\sim$ 1.604 & 44.68$\times$ $\sim$ 44775.14$\times$ & 2169.64$\times$ \\ 
\hline
\textbf{G6} &
0.063 $\sim$ 1263.6 & 0.050 $\sim$ 1233.1 & 0.89$\times$ $\sim$ 45.44$\times$ & 4.21$\times$ & 0.000 $\sim$ 42.312 & 29.73$\times$ $\sim$ 745271$\times$ & 8699.55$\times$ \\ 

\Xhline{2\arrayrulewidth}
\multicolumn{8}{l}{\textbf{G1: citeseer} ($|V|$=3327, $|E|$=4676), \textbf{G2: cora} ($|V|$=2708, $|E|$=5278), \textbf{G3: pubmed} ($|V|$=19,717, $|E|$=44,327)} \\

\multicolumn{8}{l}{\textbf{G4: com-amazon} ($|V|$=334,863, $|E|$=925,872), \textbf{G5: com-dblp} ($|V|$=317,080, $|E|$=1,049,866), \textbf{G6: com-youtube} ($|V|$=1,134,890, $|E|$=2,987,624)} \\
     
    \Xhline{2\arrayrulewidth}
    \end{tabular}
    \caption{Memory comparisons among the single-stage local PPR on CPU, the multi-stage MeLoPPR on CPU, and the MeLoPPR on FPGA.}
    \label{tab:result-all}
\end{table*}

\begin{figure*}
    \centering
    \includegraphics[width=0.96\textwidth]{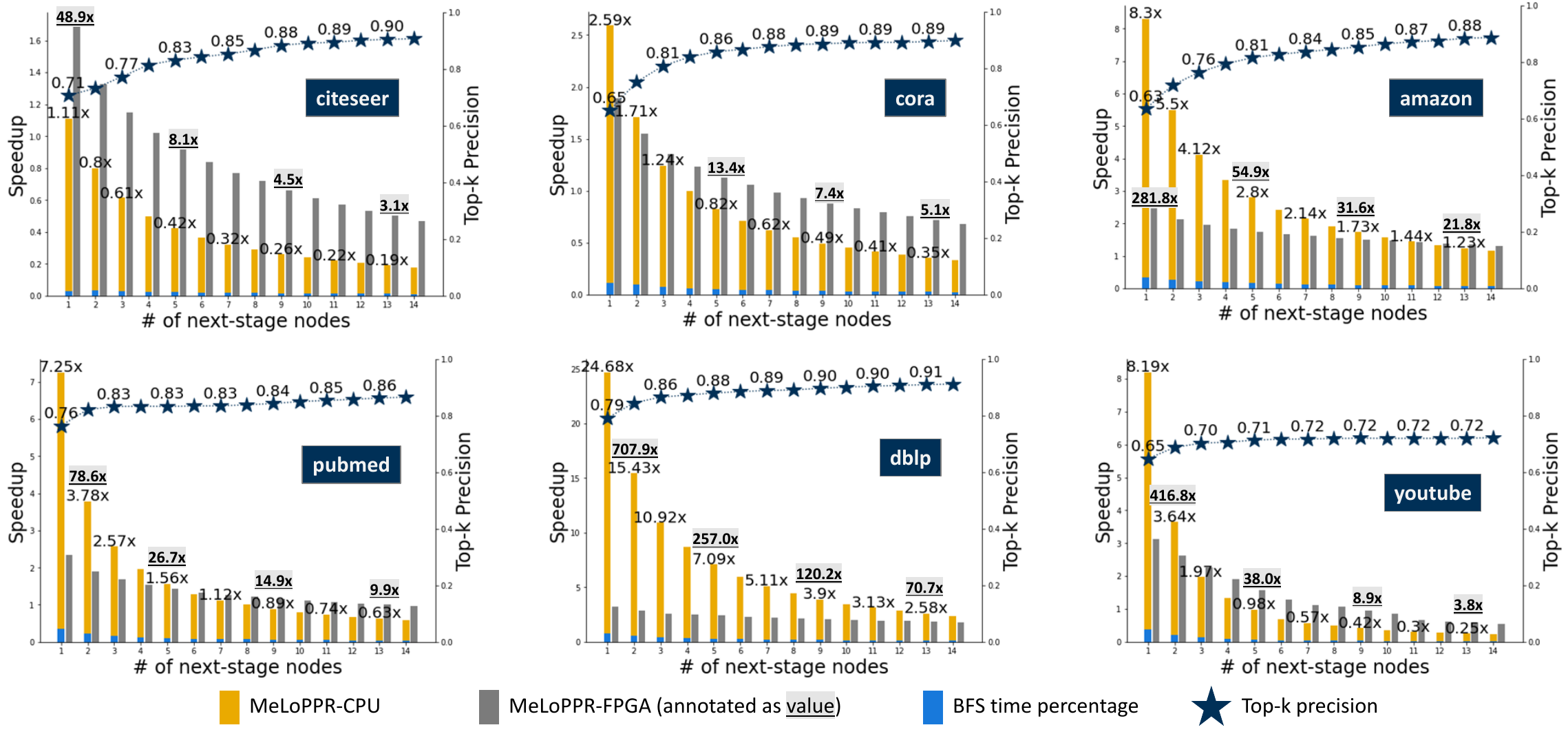}
    \caption{Precision-latency trade-offs of MeLoPPR-CPU and MeLoPPR-FPGA comparing with the baseline local PPR on CPU.}
    \label{fig:trade-off-all}
    \vspace{-8pt}
\end{figure*}

In this section, we fully evaluate MeLoPPR regarding its precision, memory efficiency, latency, and hybrid CPU+FPGA implementation.
We take six most commonly used real-world networks \cite{snapnets}, as shown in Table~\ref{tab:result-all}, where $G4$ to $G6$ are large-scale ones.
The software implementation is based on NetworkX Python library, which also serves as the comparison baseline. 
The matrix storage and matrix-vector multiplications are in compressed sparse row (CSR) format.
The CPU implementations are done on a desktop with Intel i7 core, 2.8GHz with 16 GB memory.
The FPGA implementation is on Xilinx Kintex-7 KC705 evaluation board under 100MHz clock frequency.
We let $k=200$, $L=6$, and $l_1 = l_2 = 3$ for all the experiments so that MeLoPPR contains two stages.

\vspace{-4pt}
\subsection{{FPGA Scalability Study}}\label{sec:fpga-scalability}
We conduct a scalability study to evaluate our proposed FPGA accelerator and compare it with CPU implementation. We use $G1$ as a case study and scale the parallelism $P$ from 1 to 16.
Accordingly, the number of instances of diffuser $M$ and BRAM blocks increases along with $P$.
Fig.~\ref{fig:fpga-scalability} shows the FPGA latency comparing with CPU for graph diffusion when $P$ is 1, 2, 4, 8 and 16, respectively, under 100\;MHz. It shows that improving the parallelism can effectively reduce the overall latency, over $10\times$ improvement when scaling from 1 to 16.
Meanwhile, the scheduling overhead, introduced by conflict read and write among diffusers and score tables, accounts for less than $20\%$ when $P=2$, and less than $40\%$ when $P>2$ in this experiment.
Table~\ref{tab:FPGA-resource} shows the resource utilization under different parallelism values.

\subsection{{Memory Efficiency}}
We evaluate the memory efficiency of MeLoPPR in Table~\ref{tab:result-all} by comparing with the local PPR on CPU.
For pure CPU implementation, 
the memory usage is captured by the \textit{tracemalloc} built-in module in Python.
For CPU+FPGA implementation, the memory required for FPGA is a function of the size of sub-graphs $G_{l}$ which is related to it node number $
|V(G_{l})|$ and edge number $|E(G_{l})|$. 
As shown in Fig.~\ref{fig:overall_FPGA_CPU}, for each sub-graph, there are three tables maintained in FPGA: sub-graph table ($B_g$), accumulate score table ($B_a$), and residual score table ($B_{r}$). The total FPGA memory requirement for the sub-graph in Bytes, denoted as $BRAM|_{Bytes}$, can be computed as $BRAM|_{Bytes} = B_g + B_a + B_r$ = $4\cdot ( 2 \cdot |V(G_{l})| + 2\cdot |E(G_{l})| + 2\cdot |V(G_{l})| + |V(G_{l})| )$.
Table~\ref{tab:result-all} shows that, the CPU implementation of MeLoPPR saves memory by factors of $1.51\times$ to $13.43\times$ on average, because of the significant smaller size of sub-graphs.
We observe that the denser graphs benefit from larger memory saving, such as $G3$, $G4$, and $G5$.
On FPGA, the memory requirement is $73.6\times$ to $8699\times$ smaller than CPU, because of its customized storage system that eliminates storage redundancy.

\vspace{-4pt}
\subsection{Efficiency v.s. Precision}
We demonstrate the benefits of PPR vector sparsity, which is the foundation of achieving low-latency and the precision-latency trade-off.
Fig.~\ref{fig:sparsity-and-accuracy} shows the precision results averaged from 1000 random runs on $G1$, $G2$, and $G3$.
The top figure is the precision curve when different percentages of the next-stage nodes are selected for second-stage computation. 
The smaller one contains the selection ratio from $0\%$ to $30\%$, and the larger one is zoomed from range $0\%$ to $5\%$.
The bottom figure is the distribution of normalized PPR scores in log scale to show its sparsity.
It first shows that more than $90\%$ of the nodes have near-zero PPR scores, while only less than $1\%$ nodes have large PPR scores.
Together with the top figure, it shows that if only $1\%$ of the next-stage nodes are selected, the MeLoPPR can already achieves $73.8\%$ precision (averaged from three graphs);
if $2\%$ nodes are selected, the precision is $78.1\%$, and if $3\%$ nodes are selected, the precision is $85.2\%$.
Such sparsity allows the MeLoPPR to get a satisfying precision not only with reduced memory, but also with flexible trade-offs between precision and computation latency.
Notably, the precision achieves $96.1\%$ and $96.9\%$, respectively, if $20\%$ and $30\%$ of the nodes are selected.
 
We further show the precision-latency trade-offs for all the six graphs in Fig.~\ref{fig:trade-off-all}, averaged from 500 seed nodes for each graph.
The yellow and gray bars are the speedups of MeLoPPR on CPU and FPGA comparing with the CPU baseline, the dark blue stars show the top-k precision, and the light blue bars represent the percentage of BFS on the CPU for sub-graph preparation.
It clearly shows that the precision improves and the speedup decreases while the number of computed next-stage nodes increases. For MeLoPPR-CPU, there are slowdown cases when higher precision is required (e.g. $G1$, $G2$, $G6$), while other cases (e.g. $G3$ and $G5$) indicate that MeLoPPR-CPU can achieve $1.2\times$ and $2.58\times$ speedup and reach $90\%$ precision with $6\times$ and $13.4\times$ less memory.
For MeLoPPR-FPGA, we let the parallelism $P=16$, i.e., with 16 instances of diffuser modules and 16 BRAM blocks. This is because when $P$ is larger, the BFS extraction on CPU will become the bottleneck, and increasing FPGA parallelism is no longer benefiting. 
It shows that the FPGA implementation achieves significant speedup comparing with the baseline, from $3.1\times$ to $21.8\times$, while the precision is around $90\%$.
This clearly demonstrates the necessity of using FPGAs for acceleration for PPR problems.

Through linear decomposition, MeLoPPR allows multiple next-stage nodes to be computed in parallel, which can further reduce the overall latency. We leave this for future experiments.

%% file: sec7.tex
\section{conclusion}
\label{sec:conclusion}

In this work, through software-hardware co-design, we propose a memory-efficient, low-latency personalized pagerank algorithm, namely MeLoPPR. 
MeLoPPR decomposes a single-stage PPR into multi-stage to achieve memory-efficiency, and exploits the PPR vector sparsity to achieve low-latency.
We implement the proposed MeLoPPR on a pure CPU and a hybrid CPU+FPGA platform and demonstrate a remarkable memory saving on CPU and FPGA, $6.53\times$ and $2023\times$, respectively.
We also demonstrate the flexible trade-off between latency and precision, where the FPGA implementation can achieve $3.1\times$ to $21.8\times$ speedup approximately $90\%$ precision.
This is the first work that focuses on local PPR algorithm with a tight memory budget, which opens the opportunities for low-latency parallel PPR computation on edge devices.